\documentclass[useAMS,usenatbib]{mn2e}

\usepackage{amsmath}
\usepackage{amssymb}
\usepackage[dvipsnames]{xcolor}
\usepackage{graphicx}
\usepackage{graphics}
\usepackage{hyperref}
\usepackage{comment}

\hypersetup{
     colorlinks   = true,
     citecolor    = blue
}

\def\apj{{ ApJ}}
\def\apjl{{ApJL}}

\def\aap{{ A\&A}}

\def\mnras{{ MNRAS}}
\def\araa{{ ARA\&A}}

\def\nat {{ Nature}}

\def\prd{{ Phys. Rev. D}}
\def\jcap{{ JCAP}}

\def\mr{\mathrm}
\def\d{\mr{d}}
\def\b{\mathbf}

\def\dmex{\mr{DM_{\rm ex}}}

\defcitealias{chime1}{C1}
\defcitealias{chime2}{C2}
\defcitealias{chime3}{C3}
\defcitealias{chime4}{C4}

\newcommand{\lara}[1]{{\langle#1\rangle}}

\newcommand{\myemail}{wenbinlu@caltech.edu}

\title[CHIME repeating FRBs]{Implications of CHIME repeating fast radio bursts}
\author[Lu, Piro, \& Waxman]
  {Wenbin Lu$^1$\thanks{\myemail}, Anthony L. Piro$^2$, and Eli Waxman$^3$\\
  $^1$Theoretical Astrophysics, Walter Burke Institute for Theoretical Physics, Mail Code
  350-17, Caltech, Pasadena, CA 91125, USA\\
  $^2$The Observatories of the Carnegie Institution for Science, 813 Santa Barbara Street, Pasadena, CA 91101, USA \\
  $^3$Department of Particle Physics \& Astrophysics, Weizmann Institute of Science, Rehovot 76100, Israel}

\begin{document}
\label{firstpage}
\maketitle

\begin{abstract}
CHIME has now detected 18 repeating fast radio bursts (FRBs). We explore what can be learned about the energy distribution and activity level of the repeaters by fitting realistic FRB population models to the data.
For a power-law energy distribution $\d N/\d E\propto E^{-\gamma}$ for the repeating bursts, there is a critical index $\gamma_{\rm crit}$ that controls whether the dispersion measure (DM, a proxy for source distance) distribution of repeaters is bottom or top-heavy. We find $\gamma_{\rm crit}=7/4$ for Poisson wait-time distribution of repeaters in Euclidean space and further demonstrate how it is affected by temporal clustering of repetitions and cosmological effects.
It is especially interesting that two of the CHIME repeaters (FRB 181017 and 190417) have large $\rm DM\sim 10^3\rm\, pc\, cm^{-3}$. These can be understood if: (i) the energy distribution is shallow $\gamma=1.7^{+0.3}_{-0.1}$ ($68\%$ confidence) or (ii) a small fraction of sources are extremely active. In the second scenario, these two high-DM sources should be repeating more than 100 times more frequently than FRB~121102, and the energy index is constrained to be $\gamma = 1.9^{+0.3}_{-0.2}$ ($68\%$ confidence). In either case, this $\gamma$ is consistent with the energy dependence of the non-repeating ASKAP sample, which suggests that they are drawn from the same population. Finally, our model predicts how the CHIME repeating fraction should decrease with redshift, and this can be compared with observations to infer the distribution of activity level in the whole population.

\end{abstract}


\section{Introduction}
Since the discovery of the first repeater FRB 121102 \citep{2016Natur.531..202S}, it is clear that a significant fraction of fast radio bursts (FRBs) are from non-cataclysmic sources. This is supported by the detection of 18 more repeaters by the Canadian Hydrogen Intensity Mapping Experiment \citep[][hereafter \citetalias{chime1}, \citetalias{chime2}, \citetalias{chime3}, \citetalias{chime4}]{chime1, chime2, chime3, chime4}. A common property shared by these repeaters (at least the ones with $>3$ bursts) is that fainter bursts are more common than brighter ones \citep[e.g., ][\citetalias{chime1}-\citetalias{chime4}]{2016ApJ...833..177S, 2017ApJ...850...76L, 2019ApJ...877L..19G, 2019ApJ...887L..30K, 2019arXiv191212217O}. For FRB 121102, the energy distribution of bursts can be modeled by a power-law $\d N/\d E\propto E^{-\gamma}$, but the index $\gamma$ is debated due to the lack of a homogeneously selected sample that spans a sufficiently wide range of burst energy\footnote{This is further complicated by the well-known temporal clustering of repetitions, which means a large number of observing sessions are needed to statistically analyze the rate of the brightest bursts.}. Such a distribution means that it is generally more difficult to detect a repeating source if it is located at a larger distance. For instance, if FRB 121102 were at redshift $z=1$, then the CHIME fluence threshold of a few $\rm Jy\,ms$ corresponds to an energy threshold of $\sim 10^{32}\rm\, erg\,Hz^{-1}$. However, none of the observed bursts from FRB 121102 are sufficiently bright to exceed this threshold. We see that the observed rate of such bright bursts should depend on $\gamma$ and that this should control whether repeaters can be identified at high redshift.

In this paper, we construct realistic repeating FRB population models and compare the observed dispersion measure (DM) distribution of CHIME repeaters to the model predictions in order to constrain $\gamma$. We take into account non-Poissonian wait-time clustering \citep{2018MNRAS.475.5109O}, the large number of CHIME observing sessions, redshift evolution of source densities, FRB frequency spectrum, and the stochastic DM contributions from the host galaxy and the inhomogeneous intergalactic medium (IGM). Despite the possible existence of cataclysmic FRBs, here we focus on the intrinsically repeating population and aim to test whether they are consistent with observations of both singly detected sources (referred to as ``apparent non-repeaters") and those detected multiple times (which we call ``repeaters").

The paper is organized as follows. In \S\ref{sec:motivation}, we provide a simple toy model that demonstrates the logic of our method and shows how $\gamma$ can be physically constrained by the DM distribution of repeaters. In \S\ref{sec:application}, we follow similar arguments to construct two different population models, one of which assumes that all repeaters have the same energy distribution function and the other one assumes a broad range of activity levels (some are more active than others). The results for these two cases are presented in \S\ref{sec:single} and \S\ref{sec:broad}, respectively. We discuss the implications of the inferred value of $\gamma$ and the link between repeaters and the apparent non-repeaters in \S\ref{sec:implication}. Then in \S\ref{sec:frep}, we show that, in the future, the CHIME repeating fraction can be used to infer the distribution of activity levels among different sources. Various caveats in our modeling are discussed in \S\ref{sec:caveats}. We provide a summary of our findings in \S\ref{sec:summary}. We adopt the latest Planck $\Lambda$CDM cosmology \citep{Planck16}.

\section{Motivation}\label{sec:motivation}

Generally, a very steep (or soft) energy distribution means that luminous bursts are extremely rare and hence most repeaters should be found in the local Universe. Conversely, a very shallow (or hard) energy distribution means luminous bursts are common and hence should be detected far away. This can be illustrated by the following simple model. For the case of a Euclidean universe (without redshift factor), a survey with fluence threshold $F_{\rm th}$ will only be able to detect bursts above the energy threshold $E_{\rm th} = 4\pi D^2 F_{\rm th}$ where $D$ is the distance. Hence, for a given source, the cumulative detection rate is $\dot{N}(>E_{\rm th}) \propto E_{\rm th}^{1-\gamma}\propto D^{2-2\gamma}$. If we take the Poisson waiting time distribution in the limit of small detection probability (appropriate at sufficiently large distances $D$), then the probability of detecting two bursts from the same source is $P_{\rm rep} \propto [\dot{N}(>E_{\rm th})]^2 \propto D^{4-4\gamma}$. The cumulative distance distribution of detected repeaters from a given survey is $N_{\rm rep}(<D)\propto D^3 P_{\rm rep}(D)\propto D^{7-4\gamma}$. For this particular example, we find that most repeaters will be detected at large distances if $\gamma < 7/4$, and for very steep energy distribution $\gamma\gg 7/4$, nearly all detected repeaters should be nearby. We also see that the distance distribution depends very strongly on $\gamma$, which means that this index can be effectively constrained by the DM (an indicator of the source distance) distribution of even a small sample of detected repeaters. We note that the cumulative distance distribution for singly detected sources is $N_{\rm sig}(<D)\propto D^{5-2\gamma}$ \citep[e.g.,][]{macquart18_rate_counts}, which corresponds to a critical index of 2.5 (instead of the much shallower value of 7/4 for repeaters).

\section{Model Description}
\label{sec:application}

We consider an idealized survey covering a certain patch of the sky with solid angle $\Omega$ in $n$ independent observing sessions, each of which has duration $T$. The fluence-complete threshold of the survey is denoted as $F_{\rm th}$. For the realistic CHIME/FRB survey, the fluence thresholds at the locations of different repeaters range between 2.5 and 6$\,\rm Jy\,ms$ (see Table 1's of \citetalias{chime2, chime3}, and we only consider the more sensitive upper transits), so we take $F_{\rm th}\simeq 4\rm\,Jy\,ms$ as a representitive value. The total exposure time for difference sources range from about 20 to 60 hours, and each transit lasts for about 10 minutes, so we take $T\simeq 0.2\rm\,hr$ and $n\simeq 150$. We have tested that our conclusions are insensitive to order-unity variations of $F_{\rm th}$, $n$ and $T$.

We model the distribution of time intervals between adjacent bursts from a repeating source using the Weibull distribution \citep{2018MNRAS.475.5109O}, which is parameterized by the mean repeating rate $r$ as well as the degree of wait-time clustering $k$. When $k<1$, the distribution describes that the bursts are clustered in that small intervals are favored compared to the Poissonian case and that the presence of one burst makes the detection of of an additional burst in the near future more likely. The $k=1$ case recovers the Poisson distribution. When $k>1$, the Weibull distribution can approximately describe skewed or symmetric normal distributions (the signal becomes nearly periodic in the limit of $k\gg 1$). We find our results to be insensitive to $k$ for $k\gtrsim 1$, because the detection rate is simply proportional to the exposure time.

The Weibull distribution is chosen because it provides the simplest extension of the Poisson distribution by including wait-time clustering, but other choices are also possible \citep[see][]{connor16}. Our goal is to marginalize over $k$ and provide physical constraints on the energy distribution function --- the mean repeating rate above certain burst energy $r(>E)$ as a function of $E$. In Appendix \S\ref{sec:single_repeater}, we provide a detailed description of the Weibull distribution, and the main results are the probability for at least one detection $P_{\rm det}[r(>E), k, n, T]$ for a given source and the probability for repeating detection $P_{\rm rep}[r(>E), k, n, T]$, as given by equations~(\ref{eq:6}) and (\ref{eq:7}), respectively.

In the following subsections, we explore two main models for the FRB sources. In the first case in \S\ref{sec:single}, we assume a single population of FRBs that obey the same energy dependence for their rates. In the second case in \S\ref{sec:broad}, we relax this assumption to allow a broad range of rates, which represents a situation where some FRB sources are more active than others.

\subsection{Single Population}
\label{sec:single}
We take the burst rate above a specific energy $E$ (in units of $\rm\,erg\,Hz^{-1}$) to be in the Schechter form\footnote{The Schechter form including a maximum energy guarantees the convergence of the integrated rate of energy release $\int \d E (\d r/\d E) E$ when $\gamma<2$. Our results are only weakly affected by the maximum energy since observations indicate $E_{\rm max}\gtrsim 10^{33.5}\rm\, erg\, Hz^{-1}$ \citep{2019ApJ...883...40L}, which corresponds to a fluence of $\gtrsim 56\rm\, Jy\, ms$ (much above the threshold we consider) at redshift of 1. There may physically be a (so-far unobserved) minimum energy $E_{\rm min}< 10^{28}\rm\, erg\,Hz^{-1}$ to satisfy energy convergence if $\gamma>2$, but the moderate-sensitivity survey ($F_{\rm th}\sim$a few $\rm Jy\rm\, ms$) we are considering is unaffected by $E_{\rm min}$. } with $\gamma > 1$,
\begin{equation}
  \label{eq:19}
    {\d r\over \d E} = {r_0\over E_{0}} 
\left(E\over E_0\right)^{-\gamma} \mr{exp}\left(-{E\over E_{\rm
         max}}\right),
\end{equation}
which means that fainter bursts occur more frequently than brighter ones and that there is a maximum energy $E_{\rm max}$ above which the rate cuts off exponentially. The rate of individual sources is normalized at energy $E_0 = 10^{30}\rm\,erg\,Hz^{-1}$, and $r_0$ is in units of $\rm hr^{-1}$.

The comoving number density of FRB sources as a function of redshift $z$ (for $z\lesssim 1$) takes the form
\begin{equation}
  \label{eq:29}
  n_*(z) = n_0 (1 + z)^{\beta},
\end{equation}
where $n_0$ (in $\rm Gpc^{-3}$) is the local number density and $\beta$ describes the cosmological evolution of FRB sources \citep[e.g., $\beta\simeq 2.7$ if FRB sources trace cosmic star formation history,][]{2014ARA&A..52..415M}. For a given source at redshift $z$ and luminosity distance $D_{\rm L}(z)$, the survey is only sensitive to bursts above energy
\begin{equation}
  \label{eq:47}
  E_{\rm th} = 4\pi D_{\rm L}^2 F_{\rm th} (1+z)^{\alpha-1},
\end{equation}
where we have used k-correction by adopting an intrinsic spectrum of $E_\nu\propto \nu^{-\alpha}$. We use $\alpha=1.5$ as motivated by statistical studies of the ASKAP sample \citep{2019ApJ...872L..19M}. Our results are insensitive to the spectral slope, as long as it is not extremely steep, $\alpha\lesssim 3$ \citep[e.g.,][]{2018ApJ...867L..12S}. We integrate equation (\ref{eq:19}) to obtain the rate of events above $E_{\rm th}$,
\begin{equation}
  \label{eq:42}
  \begin{split}
        r(>E_{\rm th}) &= r_0 \int_{E_{\rm th}}^\infty {\d E\over E_0}\,
  \left(E\over E_0\right)^{-\gamma} \mr{exp}\left(-{E\over E_{\rm 
         max}}\right)\\
         &= r_0 \int_{x_{\rm th}}^\infty \d x\, x^{-\gamma}
   \mr{exp} \left(-xE_0\over E_{\rm max}\right),
  \end{split}
\end{equation}
where we have used $x\equiv E/E_0$ and $x_{\rm th} \equiv E_{\rm th}/E_0$. For CHIME bursts at relatively low redshifts $z\lesssim 1$, the threshold energy is $E_{\rm th}\lesssim 10^{32}\rm\,erg\,Hz^{-1}$, which is much below the maximum energy $E_{\rm max}$ since various surveys including ASKAP, CHIME, UTMOST, and Parkes have seen bursts with $E\gtrsim 10^{33.5}\rm\,erg\, Hz^{-1}$ \citep[see e.g., Figure 1 of ][]{2019ApJ...883...40L}. Therefore, the exponential factor in equation~(\ref{eq:42}) can simply be set to unity and we obtain
\begin{equation}
  \label{eq:44}
  r(>E_{\rm th}) \approx {r_0\over \gamma -1} \left( E_{\rm th}\over
    E_0\right)^{1-\gamma},\ \ \mbox{for}\ E_{\rm th}/E_{\rm max}\ll 1,
\end{equation}
for the cumulative rate.

\begin{figure*}
\centering
\includegraphics[width=0.7\textwidth]{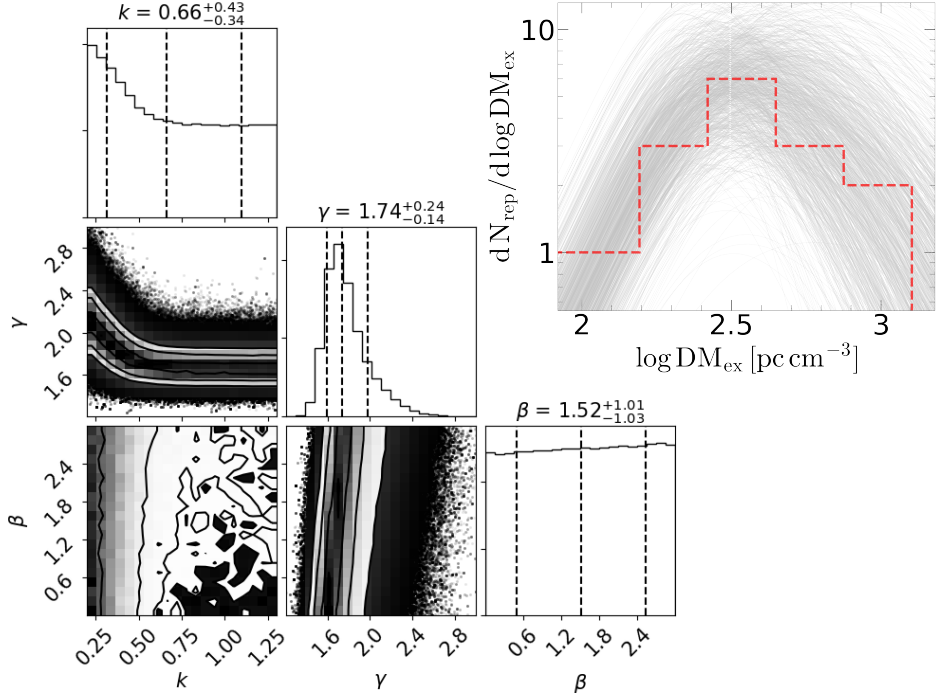}
\caption{The posterior distribution for $\b{p}=(k, \gamma, \beta)$ assuming a single population with $r_0=0.1\rm\, hr^{-1}$, as constrained by the $\rm DM_{\rm ex}$ distribution of the CHIME sample of repeating FRBs. The three vertical dashed lines in the marginal distributions marks where the cumulative density function (CDF) equals to $0.16,\ 0.5,\ 0.84$ (from left to right). The titles show the median (CDF$=0.5$) and the error range at 68\% ($1\sigma$) confidence level. The upper right plot shows the predicted $\dmex$ distributions (silver lines) for $10^3$ cases randomly drawn from the posterior distribution, as compared to observations (red dashed histogram). This plot was generated with the public code $\mathtt{corner.py}$ by \citet{corner}.
}
\label{fig:MCMC_rep_3D}
\end{figure*}

For a given source, the probability for at least one detection is $P_{\rm det}[r(>E_{\rm th}), k, n, T]$ and the probability for repeating detection is $P_{\rm rep}[r(>E_{\rm th}), k, n, T]$, as given by equations~(\ref{eq:6}) and (\ref{eq:7}), respectively. If we let $z_{\rm c}$ be the critical redshift below which more than half of the detected sources have two or more detection, which is found by setting $r(>E_{\rm th}) nT\simeq 1$ (corresponding to $P_{\rm det}\sim 1$ and $P_{\rm rep}\sim 0.5$, see Figure \ref{fig:mul_obs}), we estimate
\begin{equation}
  \label{eq:43}
  z_{\rm c}\simeq 0.1 \left[r_0nT \over (\gamma -1) \right]^{1\over 2(\gamma-1)}. 
\end{equation}
For $r_0 = 0.1\rm\, hr^{-1}$ and $\gamma= 1.8$, we find $z_{\rm c}\simeq 0.2$, which roughly agrees with the single detection of FRB 121102 (z = 0.19) by CHIME (and repetition is expected in the near future). This critical redshift can be constrained in the future by comparing the redshift distribution of single and repeating bursts in the CHIME sample (see \S\ref{sec:frep}).

Using the above framework, we next compute the expected distribution of repeating sources to compare with the observed sample of CHIME repeaters, which is plotted in the upper right panel of Figure~\ref{fig:MCMC_rep_3D}. For each source, we calculate the $\rm DM_{\rm ex}$ beyond the Milky Way by taking the observed DM and subtracting the contributions from the Galactic interstellar medium \citep[ISM,][]{2017ApJ...835...29Y} and an additional $30\rm\, pc\, cm^{-3}$ from the Galactic halo \citep{2019MNRAS.485..648P, 2020ApJ...888..105Y, 2020arXiv200111105K}. Despite the small number of bursts, we find that a significant fraction of repeaters have $\rm DM_{\rm ex}\sim 10^3\rm\, pc\, cm^{-3}$. We discuss the implications of this further below.

The cumulative distribution of repeating sources as a function of redshift is given by
\begin{equation}
  \label{eq:41}
  \begin{split}
    N_{\rm rep}(<z) = &\, {n_0\Omega\over 4\pi} \int^z \d z {\d V\over \d z} (1 + z)^{\beta} \\
    &\,\times P_{\rm rep}[r(>E_{\rm th}), k, n, T/(1+z)].
  \end{split}
\end{equation}
For a given set of parameters $\b{p} = (k, \gamma, \beta)$, we compute $N_{\rm rep}(<z)$ and then convert it into $\d N /\rm \d log\, DM_{\rm ex}$, where $\rm DM_{\rm ex}$ is the dispersion measure to the source minus the Milky Way contribution\footnote{The conversion between $N_{\rm rep}(<z)$ and $\d N /\rm \d log DM_{\rm ex}$ is discussed in detail in Appendix \ref{sec:DM_z} where we take into account the noisy DM contributions from the host galaxy and IGM.}.
We then normalize it by the expectation number of sources $\lambda_{\rm peak}$ in the peak bin near $\rm log\, DM_{\rm ex} [pc\,cm^{-3}] \simeq 2.5$. In Monte Carlo simulations, we randomly draw the expectation value $\lambda_{\rm peak}$ from the Poisson PDF $\d P/\d \lambda_{\rm peak} = \lambda_{\rm peak}^{n_{\rm peak}}\mr{exp}(-\lambda_{\rm peak})/n_{\rm peak}!$ based on the detected number\footnote{The number of sources in the peak bin near $\rm log\, DM_{\rm ex} [pc\,cm^{-3}] \simeq 2.5$ is between 4 and 7 (depending on the bin size), whose Poisson error is much less than the bin at $\rm log\, DM_{\rm ex}\simeq 3$ with only 2 sources. The overall uncertainty is dominated by where the error is the largest. This has been confirmed by varying $n_{\rm peak}$ between 4 and 7, and the resulting difference is smaller than the statistical error. We also find that directly using $n_{\rm peak}$ instead of $\lambda_{\rm peak}$ only makes a small difference compared to the overall statistical error.} $n_{\rm peak}\simeq 6$. Thus, we obtain the number of expected detections $n_{\rm exp}$ in the bin at $\rm log\, DM_{\rm ex}\simeq 3$, which is then compared with the detected number of bursts $n_{\rm obs}=2$ in this bin. Since the number of sources in the field of view is large, the number of detections is time independent and the likelihood function for this set of parameters is Poissonian,
\begin{equation}
    L(\b{p}) = n_{\rm exp}^{n_{\rm obs}} \mr{e}^{-n_{\rm exp}}/n_{\rm obs}!,
\end{equation}
which is then used to calculate the posterior distribution of the parameters $\b{p}$ using the Bayesian theorem. We use the following priors: $0.2<k<1.3$, $1.2<\gamma<3$, $0<\beta<3$. The lower limit of $k$ is motivated by the studies of FRB 121102 by \citet{2018MNRAS.475.5109O} and by the fact that the detections of repetitions are often (but not always) spread over multiple observing sessions \citep[\citetalias{chime1}-\citetalias{chime4},][]{2019ApJ...887L..30K, 2019arXiv191212217O}. The motivation for the prior on redshift evolution is that the source number density is assumed to be somewhere between non-evolving ($\beta=0$) and tracing star-formation history ($\beta\simeq 3$). The final result of the marginalized PDF for $\gamma$ depends weakly on the $\beta$ prior.

\begin{figure*}
\centering
\includegraphics[width=0.95\textwidth]{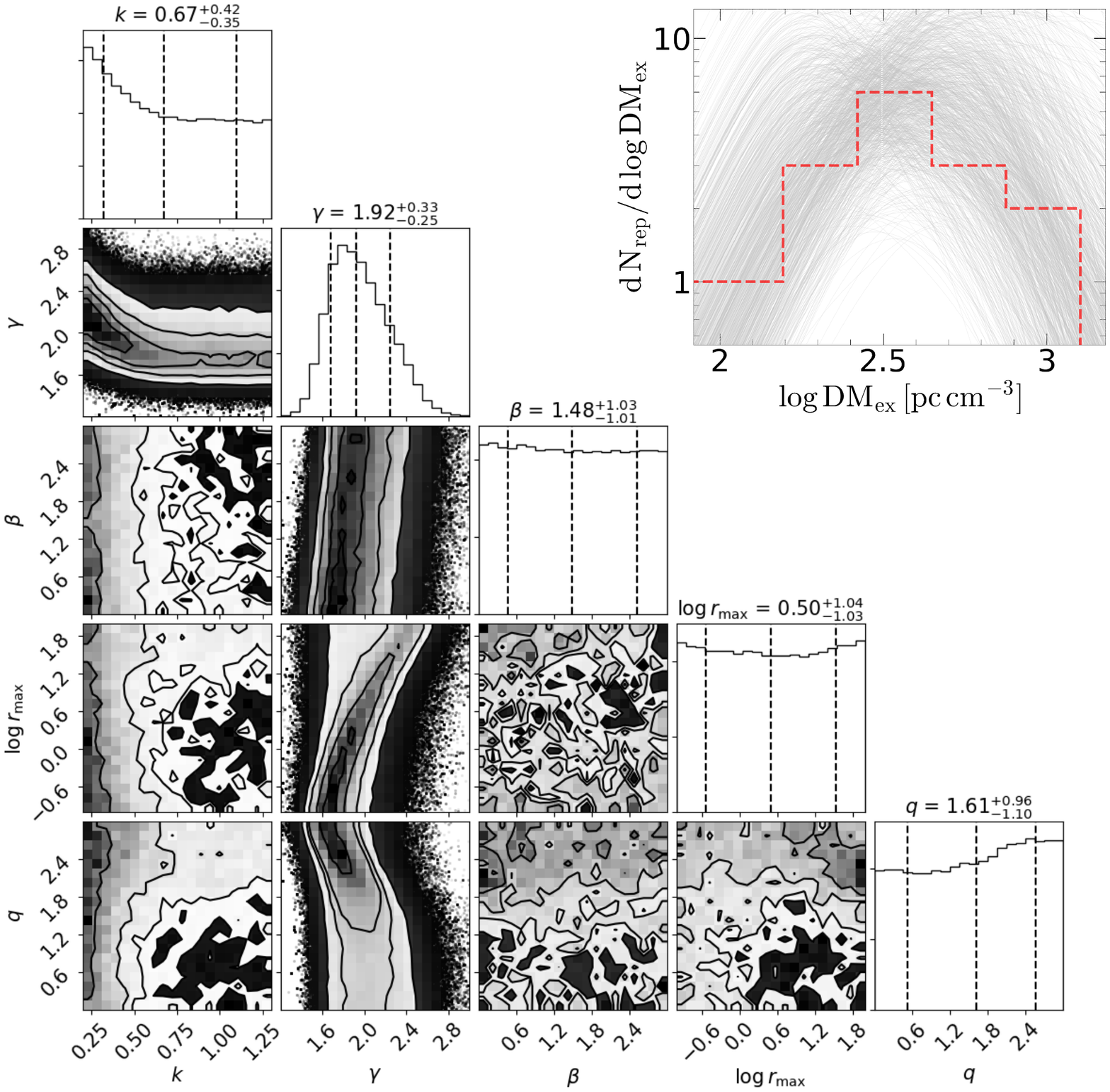}
\caption{The posterior distribution for $(k,\, \gamma,\, \beta,\, \mr{log}\,r_{\rm max} [\mr{hr^{-1}}],\, q)$ for the case of a broad distribution of $r_0$. The titles in the marginal distributions show the median and the error range at 68\% ($1\sigma$) confidence level. Note that $\gamma\gtrsim 2$ is only allowed if $r_{\rm max}\gtrsim 10\rm\, hr^{-1}$, i.e. the most extreme repeaters are more than $10^2$ times more active than FRB 121102. The upper right plot shows the predicted $\dmex$ distributions (silver lines) for $10^3$ cases randomly drawn from the posterior distribution, as compared to observations (the same as in Figure~\ref{fig:MCMC_rep_3D}).
}
\label{fig:MCMC_rep_5D}
\end{figure*}

The posterior distribution for $\b{p} = (k, \gamma, \beta)$ is sampled by a Markov-Chain Monte Carlo (MCMC) simulation and is shown in Figure \ref{fig:MCMC_rep_3D}. We constrain the power-law index for the energy distribution to be $1.6\lesssim \gamma \lesssim 2$ (68\% confidence interval), whereas the two other parameters $k$ (waiting time clustering) and $\beta$ (redshift evolution) are essentially unconstrained due to small number statistics. Note that the constraints given by our likelihood function are conservative because we only make use of the detected number of repeaters in two bins at $\rm log\, DM_{\rm ex} [pc\,cm^{-3}]\simeq 2.5$ and $3$ (to minimize possible systematic errors due to selection biases). Our method can be generalized to include the full $\dmex$ distribution when more bursts are available and possible selection biases are reasonably understood.

For the purpose of gaining analytical insight, in the $z\gg z_{\rm c}$ limit, we have $r(>E_{\rm th}) nT\ll 1$ and the repeating probability of each source is roughly given by equation~(\ref{eq:14}). For the simplest case of Euclidean universe (without redshift factor) and Poisson distribution ($k=1$), the cumulative distance ($D$) distribution of repeaters is given by
\begin{equation}
  \label{eq:46}
 \begin{split}
   N_{\rm rep}(<D) &\simeq {n_*\Omega (nT)^2\over 2} \int^D\d D\, D^2
  [r(>E_{\rm th})]^2\\
  &\simeq {n_*\Omega (r_0nT)^2\over 2 (\gamma-1)^2} 
    \int^D\d D\, D^2 \left(E_{\rm th} \over
    E_0\right)^{2-2\gamma},
 \end{split}
\end{equation}
which means $N_{\rm rep}(<D)\propto \d N_{\rm rep}/\d \mr{log}\,D \propto D^{7-4\gamma}$. The $\rm DM_{\rm ex}$ distribution in Figure~\ref{fig:MCMC_rep_3D} is not much steeper than $\sim D^{-1}$, implying $\gamma\lesssim 2$. Properly including comoving volume in a $\rm \Lambda$CDM Universe will predict less high-$z$ repeaters and strengthen the upper limit on $\gamma$. On the other hand, for strong waiting time clustering $k\lesssim 0.5$, we predict more high-$z$ repeaters because $P_{\rm rep}\propto r^{1+k}$ instead of $r^2$ so the larger $\gamma$ is allowed.

\subsection{Broad Distribution of $r_0$}
\label{sec:broad}
We now relax the assumption of a single population of FRB sources and consider a broad distribution of $r_0$. Such a scenario is applicable if some sources are more active than others \citep[as suggested by the non-detection of repeaters by ASKAP and follow-ups, e.g.,][]{2019arXiv191207847J}. We adopt a power-law distribution as follows
\begin{equation}\label{eq:dnstar_dr0}
    {\d n_*\over \d r_0} = {n_*\over r_{\rm max}} {q-1\over R^{q-1} - 1} (r_0/r_{\rm max})^{-q},
\end{equation}
where $n_*$ is the total number density, $r_{\rm max}$ is the maximum repeating rate normalization, $R = r_{\rm max}/r_{\rm min}$ is the ratio between the maximum and minimum repeating rates, and $q$ is a power-law index. Currently, it is unclear whether most sources are near the most active end $(r_0\sim r_{\rm max},\ q<1)$ or the least active end $(r_0\sim r_{\rm max}/R,\ q > 1)$. A possible physical scenario is the following. For each source, the repeating rate drops as a power-law function of age $r_0\propto t^{-p}$ and, if the source birth rate per unit volume is constant, the number of sources is proportional to the age $t\propto r_0^{-1/p}$, so we obtain $\d n_*/\d r_0\propto r_0^{-1-1/p}$ or $q = 1 + 1/p > 1$. 

The cumulative redshift distribution of repeaters in equation~(\ref{eq:41}) is now modified to
\begin{equation}
\begin{split}
    N_{\rm rep}& (<z) = {\Omega\over 4\pi} \int^z_0 \d z {\d V\over \d z} (1 + z)^{\beta} \int_{r_{\rm max}/R}^{r_{\rm max}} \d r_0 {\d n_*\over \d r_0} \\
    & \times P_{\rm rep}[r(>E_{\rm th}), k, n, T/(1+z)].\\
\end{split}
\end{equation}
A broad distribution of $r_0$ strongly impacts the redshift distribution of repeaters at $z\ll 1$, because the critical redshift $z_{\rm c}$, at which $P_{\rm rep}\sim 0.5$, depends on $r_0$ through equation~(\ref{eq:43}). However, the shape of the redshift distribution at $z\gg z_{\rm c}(r_{\rm max})$ is not affected, because $P_{\rm rep}\propto r^{1+k}\propto r_0^{1+k} E_{\rm th}^{(1-\gamma)(1+k)}$ and hence the $\int \d r_0$ integral separates from the $\int \d z$ integral. If the most active repeaters are similar to FRB 121102, i.e., $r_{\rm max} \sim 0.1\rm\,hr^{-1}$ \citep{2017ApJ...850...76L, 2019MNRAS.486.5934J}, which corresponds to $z_{\rm c}(r_{\rm max}=0.1\rm\, hr^{-1}) \sim 0.2$, then $z\gg z_{\rm c}(r_{\rm max})$ roughly holds for the majority of CHIME repeaters with $\dmex\gtrsim 300\rm\, pc\,cm^{-3}$. Thus, the constraints on $\gamma$ are similar to the single population case as discussed in \S\ref{sec:single}.

It is also possible that a small fraction of sources are extremely active such that the high-$z$ repeaters are dominated by the most active ones. We include two additional free parameters $\mr{log\,}r_{\rm max} [\mr{hr}^{-1}]$ and $q$ in the likelihood analysis, with sufficiently wide priors of $-1<\mr{log\,}r_{\rm max} [\mr{hr}^{-1}]<2$ and $0<q<3$ to account for this. The maximum $r_{\rm max}=10^2\rm \, hr^{-1}$ roughly corresponds to repeating sources that are $10^3$ times more active than FRB 121102 (although no such hyper-active sources have been identified observationally). We have tested that the final constraints on $\gamma$ are not sensitive to the ratio $R=r_{\rm max}/r_{\rm min}$, as long as it is sufficiently large (we adopt $R=10^6$ in practice).

The MCMC-sampled posterior distribution for $\b{p} = (k,\, \gamma,\, \beta,\, \mr{log\,}r_{\rm max} [\mr{hr}^{-1}],\, q)$ is shown in Figure~\ref{fig:MCMC_rep_5D}. We constrain the energy distribution of repeating bursts to be $\gamma=1.9^{+0.3}_{-0.2}$ (68\% confidence interval), whereas the other parameters are unconstrained due to small number statistics (see \S\ref{sec:frep} for a discussion on how $r_{\rm max}$ and $q$ may be constrained by the CHIME repeating fraction). It is interesting to look at the covariance between $r_{\rm max}$ and $\gamma$. As anticipated, we see that very large $r_{\rm max}\gtrsim 10\rm\, hr^{-1}$ (for sources more than $10^2$ times more active than FRB 121102) makes it possible to detect high-$\dmex$ repeaters without requiring a shallow energy distribution for each source, so the best-fit energy distribution index $\gamma$ is pushed to higher values (in this case, small $\gamma$ tend to over-produce the number of high-$\dmex$ repeaters). This can be directly tested by future monitoring of high-$z$ repeaters FRB 181017 and 190417, just to see whether they are hyper active with $r_0\gtrsim 10\rm\, hr^{-1}$.

\section{Implications}
\label{sec:discussion}

In this section, we first discuss the implications of our constraints on $\gamma$ and the link between repeaters and the apparent non-repeaters. Then, we show that the CHIME repeating fraction can be used to infer the distribution of activity levels among different sources.

\begin{figure}
\centering
\includegraphics[width=0.45\textwidth]{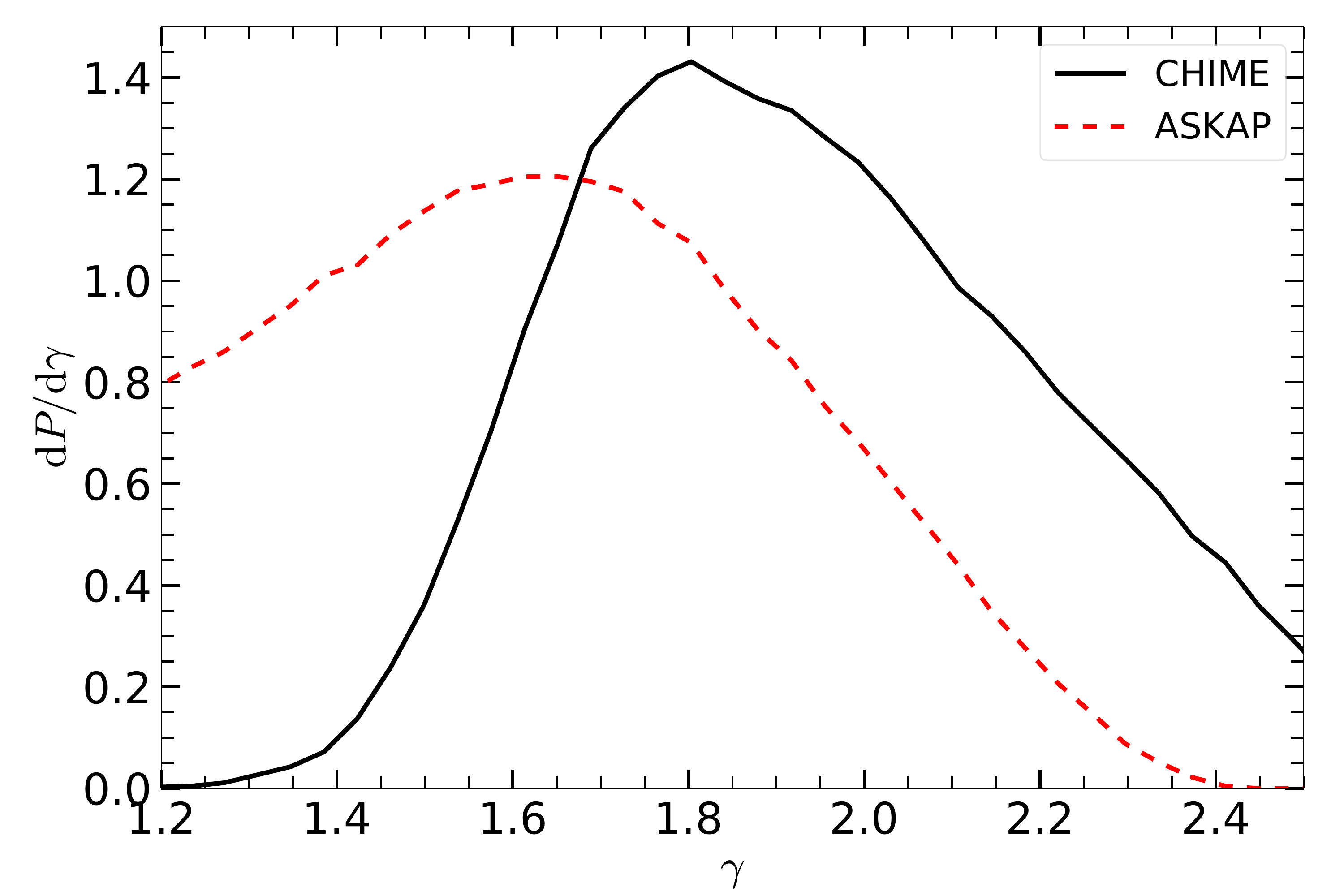}
\caption{The PDFs of the energy distribution power-law index $\gamma$ as constrained by the CHIME repeating sample (black solid line) and the ASKAP apparently non-repeating sample (red dashed line). The agreement between these two constraints suggests, although does not prove, that all FRBs are drawn from the same repeating population.
}
\label{fig:gam_joint}
\end{figure}

\subsection{Implications of $\gamma$}\label{sec:implication}
To summarize, we find that CHIME detection of high-$\dmex$ repeaters constrains $1.6\lesssim \gamma\lesssim 2$, if all sources are from the same population similar to FRB 121102 as assumed in \S\ref{sec:single}. In the more general case considered in \S\ref{sec:broad}, where some sources are allowed to be much more active than others, the constraints become $1.7\lesssim \gamma\lesssim 2.2$, slightly steeper because the detected high-$\dmex$ repeaters could simply be the most active sources (and hence do not require a shallow energy distribution for each source). If the total volumetric rate of FRBs is dominated by the sources we model (with the underlying assumption that they all repeat with the same $\gamma$), then the differential volumetric rate per energy also has the same power-law behavior
\begin{equation}
    {\d \Phi\over \d E} = \int \d r_0 {\d n_*\over \d r_0} r_0 {\d \dot{N}\over \d E} = n_* \lara{r_0} {\d \dot{N}\over \d E}\propto E^{-\gamma},
\end{equation}
where $\lara{r_0}\equiv n_*^{-1}\int \d r_0 (\d n_*/\d r_0) r_0$ is the mean rate normalization and $\Phi$ is in units of $\rm Gpc^{-3}\, yr^{-1}$. 

Independent analysis of the ASKAP sample of apparent non-repeating FRBs by the authors \citep{2019ApJ...883...40L} constrains $1.3\lesssim \gamma\lesssim 1.9$ (68\% confidence interval), which is in rough agreement with the constraints from the CHIME sample of repeaters\footnote{It is worth noting that the CRAFT survey has relatively poor threshold ($F_{\rm th}\sim 50\rm\, Jy\, ms$) and is only capable of detecting the brightest bursts \citep{2018Natur.562..386S}. The repeating rate above the threshold energy for each source $r(E_{\rm th})$ is small in that $P_{\rm det}\ll 1$ is well satisfied, so the detection probability is linearly proportional to the total observing time spent on each source, independent of the arrangement of observing runs as shown by equation~(\ref{eq:11}).}, as shown in Figure~\ref{fig:gam_joint}. This suggests, although does not prove, that all FRB sources are repeating\footnote{This is also supported by two other observations: (1) One of the ASKAP apparent non-repeaters is seen to repeat when observed with a more sensitive telescope \citep{2019ApJ...887L..30K}; (2) CHIME has only detected one burst from FRB 121102 \citep{2019ApJ...882L..18J}, so this well-known repeating source would be considered as a non-repeater if it were only detected by CHIME.} with the same $\gamma$. The errors in both constraints are dominated by the small number of sources, which will be dramatically improved by future observations.

We also note that direct measurements of $\gamma$ from individual repeaters may have large uncertainties \citep[][\citetalias{chime4}]{2017ApJ...850...76L, 2017JCAP...03..023W, 2019ApJ...877L..19G, 2019MNRAS.486.5934J}; see Table 2 of \citet{2019arXiv191212217O}. This can be understood if the monitoring time is not long enough to capture the more luminous bursts. For a given source, the Weibull waiting time distribution with $k < 0.5$ has the property that a survey either detects a large number of bursts or no burst at all \citep[see Figure 4 of][]{2018MNRAS.475.5109O}. {\it This effect leads to a bias towards larger $\gamma$ or steeper energy distribution.}

\subsection{Future constraints by repeating fraction}\label{sec:frep}
In this subsection, we show how the CHIME repeating fraction can be used to constrain the unknown parameters other than $\gamma$ and reveal the distribution of activity levels among different FRB sources (as defined in eq. \ref{eq:dnstar_dr0}). Generally, a steep activity-level distribution $q>2$ means that most sources are very inactive with $r_0\ll r_{\rm max}$ and most CHIME repeaters should be just repeating frequently enough to give repetitive detections. A very shallow distribution $q<1$ means that most sources are very active with $r_0\sim r_{\rm max}$ and hence we roughly recover the single population case as discussed in \ref{sec:single}. The intermediate region of $1 < q < 2$ is more complex in that, although most sources are not very active, the detected repeaters may or may not be dominated by the most active sources near $r_{\rm max}$.

\begin{figure}
\centering
\includegraphics[width=0.48\textwidth]{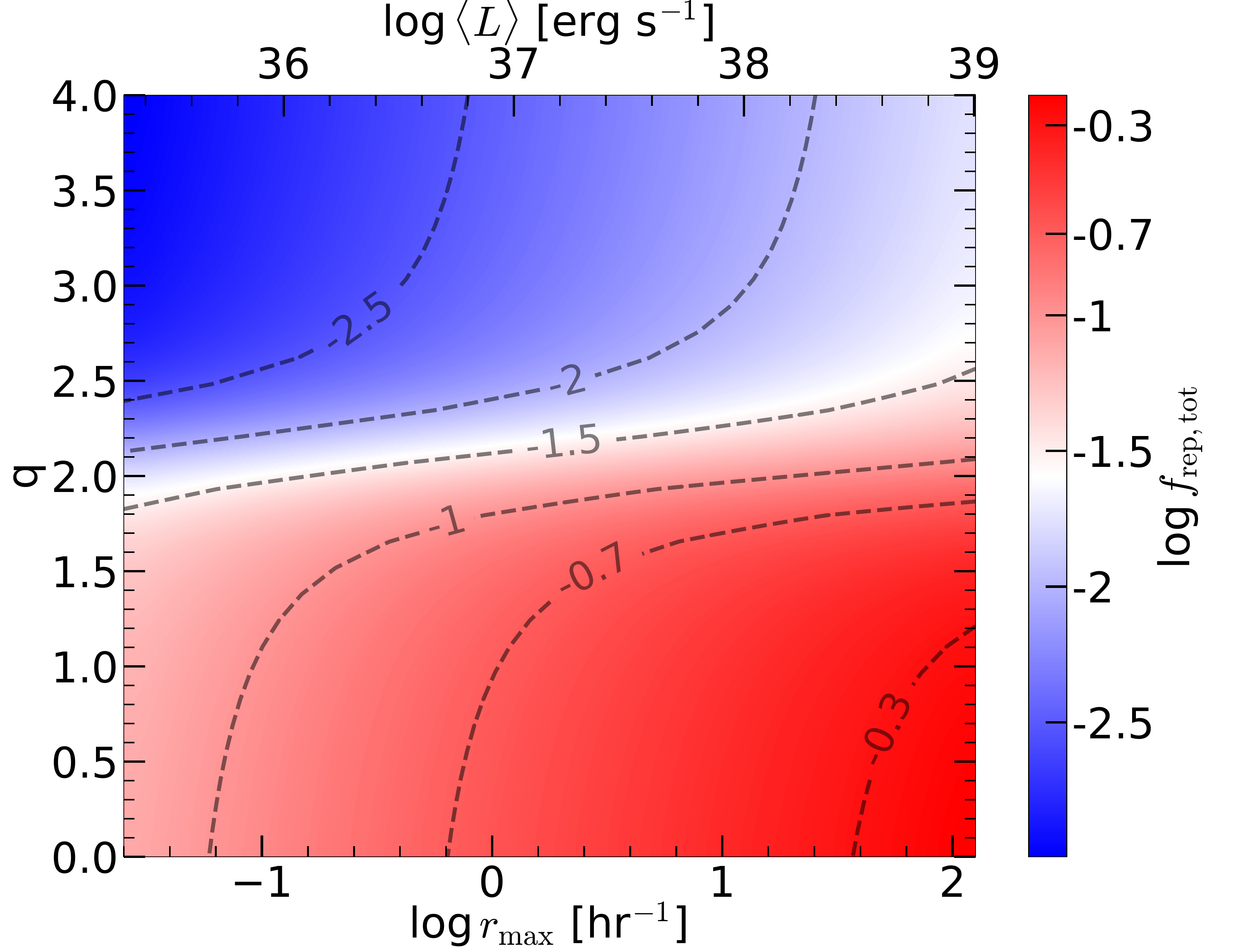}
\caption{The total repeating fraction for the CHIME survey $f_{\rm rep,tot} = N_{\rm rep}/N_{\rm det}$ within $z<2$, as a function of the maximum repeating rate normalization $r_{\rm max}$ and the activity-level distribution slope $q$ (as defined in eq. \ref{eq:dnstar_dr0}). We fix $k=1/3$ for Weibull clustering, $\gamma = 1.8$ for energy distribution, and $\beta=1.5$ for redshift evolution, and the results are qualitatively similar for other choices of these parameters. On the top axis of the plot, for the most active source at $r_{\rm max}$, we show the average luminosity by integrating over the energy distribution $\lara{L} = \int_0^\infty (\d r/\d E) E\Delta \nu \d E$, adopting a maximum energy $E_{\rm max}=10^{34}\rm\, erg\,Hz^{-1}$ \citep{2019ApJ...883...40L} and typical spectral width $\Delta \nu = 1\rm\, GHz$. Note that $\lara{L}$ is the isotropic luminosity in the radio band only, and the total energy dissipation rate of the source may be much higher, depending on the emission mechanism.
}
\label{fig:frep_tot}
\end{figure}

We calculate the total repeating fraction for the CHIME survey $f_{\rm rep,tot} = N_{\rm rep}/N_{\rm det}$ within redshift of $2$, as a function of $\mr{log}\, r_{\rm max}$ and $q$. We fix $k=1/3$ \citep[as indicated by FRB 121102,][]{2018MNRAS.475.5109O}, $\gamma = 1.8$ (as discussed \S\ref{sec:implication}), and $\beta=1.5$ for mild redshift evolution, and the results are qualitatively similar for other choices of these parameters. As shown in Figure~\ref{fig:frep_tot}, the CHIME repeating fraction increases towards larger $r_{\rm max}$ and smaller $q$. If most sources are similar to or more active than FRB 121102 ($\mr{log}\, r_{\rm max} \mr{[hr^{-1}]}\gtrsim -1$, $q<1$), then $\gtrsim 10\%$ of all CHIME sources should be repeaters. In the other extreme limit of $q>2.5$ and $\mr{log}\, r_{\rm max} \mr{[hr^{-1}]}\simeq -1$ (FRB 121102 represents the most active sources), the repeating fraction is much less than $1\%$.

The observed CHIME repeating fraction is $f_{\rm rep,tot}\sim 3\%$ \citepalias{chime3}, although this may be subjected to substantial error due to unaccounted selection biases. If we take $f_{\rm rep,tot}$ at face value, then Figure~\ref{fig:frep_tot} directly gives $1.5\lesssim q\lesssim 2.5$ with weak dependence/constraint on $r_{\rm max}$. We remark that current data allows $q=2$, which means that the overall volumetric FRB rate has comparable contributions from sources of all activity levels (the contribution from those repeating less frequently are compensated by their larger source number density).

\begin{figure}
\centering
\includegraphics[width=0.48\textwidth]{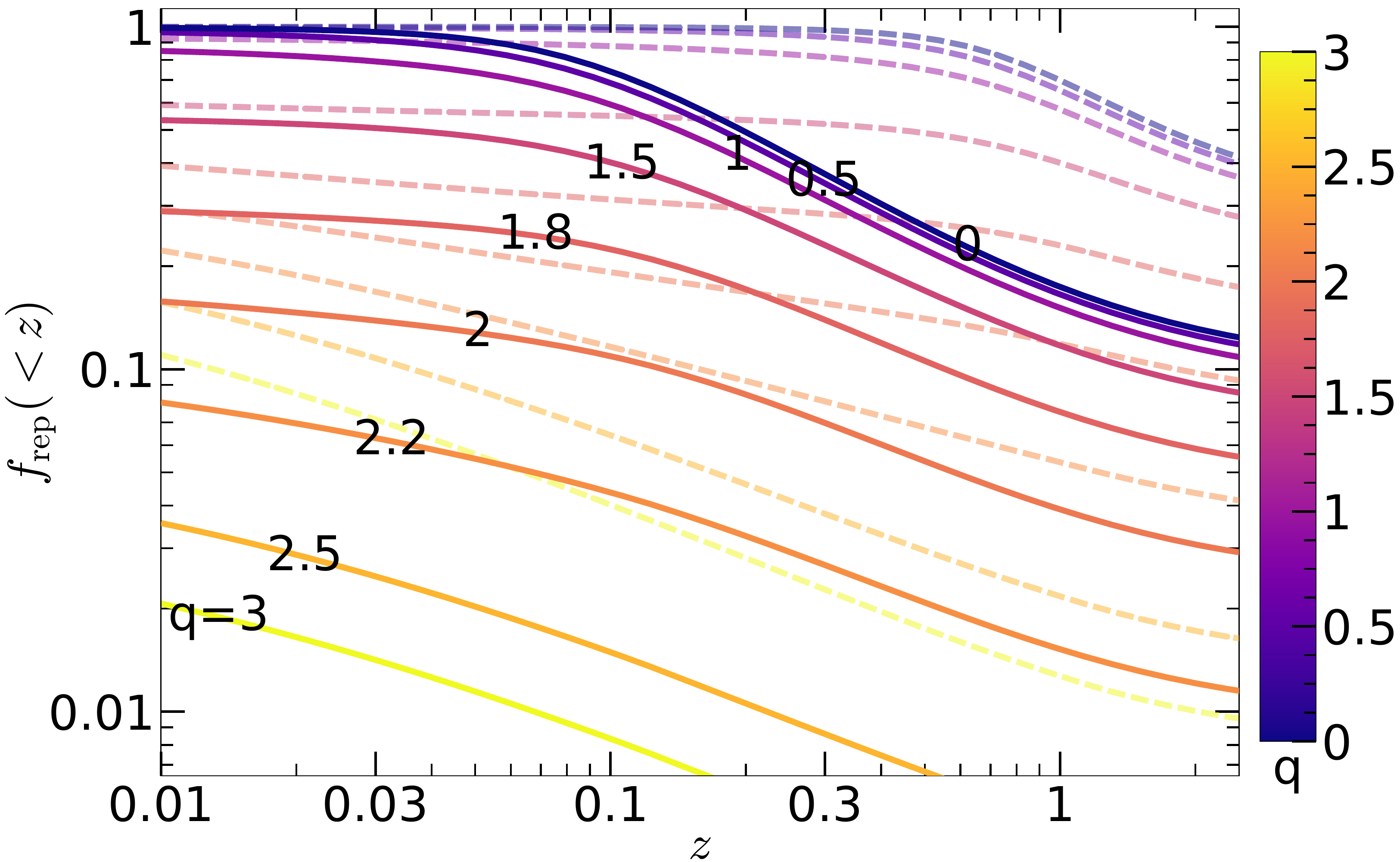}
\caption{The cumulative repeating fraction $f_{\rm rep}(<z)$ for detected sources below redshift $z$, for different maximum repeating rate normalizations $r_{\rm max}=0.1\rm\, hr^{-1}$ (solid lines) and $10\rm\, hr^{-1}$ (dashed lines), and for different the activity-level distribution slopes $q$ (color-coded). We fix $k=1/3$ for Weibull clustering, $\gamma = 1.8$ for energy distribution, and $\beta=1.5$ for redshift evolution.}
\label{fig:frep_z}
\end{figure}

More information on the population properties can be obtained by studying how the repeating fraction depends on redshift. We define the cumulative repeating fraction as $f_{\rm rep}(<z) = N_{\rm rep}(<z)/N_{\rm det}(<z)$, which is shown in Figure~\ref{fig:frep_z}, for a number of population models with different maximum repeating rate normalizations $r_{\rm max}$ and activity-level distribution slopes $q$. Again, as a representative example, we fix $k=1/3$, $\gamma = 1.8$, and $\beta=1.5$. For this example, we find that, if most sources are similar to FRB 121102 ($r_{\rm max} \sim 0.1 \mr{\,hr^{-1}}$, $q<1$), then more than $60\%$ sources at $z\lesssim 0.1$ should be repeaters. If FRB 121102 represents the most active sources ($r_{\rm max} \sim 0.1 \mr{hr^{-1}}$) but $q > 2.5$, then the repeating fraction is less than a few percent even at very low redshift $z\sim 0.01$. The overall repeating fraction is much higher for the $r_{\rm max} = 10\, \mr{hr^{-1}}$ cases.

\section{Potential Limitations}\label{sec:caveats}

In an effort to be as clear as possible about potential limitations to our analysis, we mention a number of caveats in our current modeling. These will be improved in the future with better statistics and a deeper understanding of CHIME's selection biases. 

(1) Our constraints on $\gamma$ may be subjected to CHIME selection biases if high-DM ($\sim 10^3\rm\,pc\, cm^{-3}$) bursts are more difficult to detect than the ones with the same fluence but at lower DM ($\sim 300\rm\, pc\, cm^{-3}$). The intra-channel dispersion smearing for the CHIME survey with spectral resolution $\Delta \nu$ at frequency $\nu$ is given by
\begin{equation}
    \Delta t_{\rm DM}\simeq 0.8\mr{\,ms}\, {\mr{DM} \over 10^{3} \mr{\, pc\, cm^{-3}}} {\Delta \nu\over 0.02\rm\, MHz} \left(\nu\over 600\rm\, MHz\right)^{-3},
\end{equation}
so it is likely that some narrow bursts near the lower end of the frequency band $\nu\sim 400\rm\,MHz$ are missed. It is also possible if high-DM bursts are preferentially scattering broadened. If such biases exist, then the true fraction of high-$\rm\,DM_{\rm ex}$ repeaters is larger and hence $\gamma$ should be slightly reduced (giving a shallower energy distribution).

(2) The sky positions of different sources have different exposure time and different fluence threshold. Additionally, as the CHIME beams regularly sweep across the position of a given source, the differential exposure time under the instantaneous fluence threshold may not be well modeled by a top-hat function as in our model. These complications can be included in a generalized version of our model in the future when a better understanding of the CHIME beams is available. At the current moment, since most bursts are not detected far away from the beam centers (see Figure 1's of \citetalias{chime2, chime3}), the effects of beam biases should be weak. Also, the exposure time and fluence threshold for the locations of the two highest DM sources FRBs 181017 and 190417 are close to the median in the CHIME repeater sample (see \citetalias{chime2, chime3}), so the potential biases due to non-uniform sky coverage should not be strong.

(3) Another possible complication is that the host DM contribution for FRB 181017 and 190417 may be close to $10^3\rm\, pc\, cm^{-3}$ such that they are actually located at much lower redshifts $z\ll 1$. This is possible given the uncertainties on local $(\lesssim \rm pc)$ environment of FRB progenitors and their host galaxy properties. However, the ($\sim$10) known examples of FRB host galaxies have low to modest $\rm DM_{\rm host}$ of a few 10's up to about $200\rm\, pc\,cm^{-3}$. A low local contribution is also expected for young neutron star scenarios if the observed DM is not changing appreciably over time \citep[e.g.,][]{Piro16,Piro18}. Future host localizations will test this possibility.

(4) Note that our constraints on $\gamma$ are conservative because we only make use of the detected number of repeaters in two bins at $\rm log\, DM_{\rm ex} [pc\,cm^{-3}]\sim 2.5$ and $3$. This is to minimize possible systematic errors due to CHIME selection biases in other bins (at $\dmex\lesssim 10^2$ or $\gg 10^3\rm\, pc\, cm^{-3}$). Our method can be generalized to include the full $\dmex$ distribution when more repeaters are available and possible selection biases are  understood. It is also possible to extend our model to predict the $\dmex$ distribution of the sources with more than 2 or 3 detected bursts and then compare it with observations. These additional constraints will provide information on other parameters such as the source number density $n_*$ and the maximum repeating rate $r_{\rm max}$.

(5) The possible periodicity of FRB 180916 \citepalias{chime4} cannot be captured by the Weibull distribution. If the repetition indeed has active windows followed by inactive windows, with duty cycle $f_{\rm d}$, then the effective number of observing sessions is reduced by a factor of $f_{\rm d}$ and the mean repeating rate inside the active windows is higher by a factor of $f_{\rm d}^{-1}$. We have tested that our conclusions are not sensitive to order-unity changes of the (effective) number of observing sessions.
Additionally, in Appendix \ref{sec:correlation}, we test the assumption of $n$ \textit{independent} observing sessions against the true Weibull distribution spanning the entire observing run of $n$ sessions. We adopt the former in this paper mainly to make the calculations analytically tractable. We show that the  probability of repetitive detections is
different for these two methods. However, the difference between the two can be effectively reconciled by shifting the shape parameter $k$ to a different value. Therefore, our final  results, after marginalizing over $k$, should not be strongly affected by the assumption of independent observing sessions.

\section{Summary}
\label{sec:summary}

In this work we have shown that the redshift (or $\dmex$) distribution of repeating FRBs in a given survey depends strongly on the energy distribution of repeaters and hence can be used to constrain the important property of the sources. We constructed a model for the whole FRB population based on the Weibull wait-time distribution with arbitrary clustering, properly taking into account realistic cosmological effects and that some sources may repeat more frequently than others. The model-predicted $\dmex$ distribution was then compared to the CHIME repeaters to constrain the energy distribution index $\gamma$ in a Bayesian way. Our findings are summarized as follows.

(1) Figures \ref{fig:single_obs} and \ref{fig:mul_obs} provide a sense for whether single or multiple observing sessions are expected to find repeating sources if all FRBs repeat. This can roughly be compared with future surveys and different strategies to get a better idea if they are able to rule out repetition or not.

(2) CHIME's detection of two high-$\dmex$ repeaters can be understood if either a small fraction of sources are intrinsically much more active than FRB 121102 or the energy distribution for repetitions is shallow. In the first explanation, FRBs 181017 and 190417 should be at least $\sim10^2$ times more active than FRB 121102, and this can be tested by modest amount of follow-up observations of these two sources with more sensitive telescopes than CHIME. If such extremely active sources dominate high-$\dmex$ repeaters, then the energy distribution index is constrained to be $1.7\lesssim \gamma\lesssim 2.2$. On the other hand, the second explanation gives shallower power-law index of $1.6\lesssim \gamma \lesssim 2$. This can also be tested by future monitoring of nearby repeaters.

(3) The hypothesis that all FRB sources are repeating with a universal $\gamma \sim 1.8$ is consistent with all observations\footnote{We also tried constraining other parameters (especially $r_{\rm max}$ and $q$) by imposing a prior of $\gamma=1.8$ or other similar values between 1.6 and 2. We only found that $r_{\rm max}\lesssim 10\rm\, hr^{-1}$ is favored but the current repeater data is insufficient to rule out larger $r_{\rm max}$ at high confidence.}, including the CHIME repeaters, the apparent non-repeaters found by the CRAFT survey \citep{2019ApJ...883...40L}, and FRB 121102 \citep{2017ApJ...850...76L, 2019MNRAS.486.5934J, 2019arXiv191212217O}. This power-law index is shallower than that of the Crab giant pulses \citep[$\beta\sim2.1$--3.5,][]{2012ApJ...760...64M} but consistent with magnetar X-ray bursts \citep[$\beta\sim 1.4$--2.0,][]{2015RPPh...78k6901T} and other systems displaying self-organized criticallity \citep{1986JGR....9110412K, 1987PhRvL..59..381B}. This lends indirect support to the magnetar nature of FRB progenitors (as pointed out earlier by \citealp{2016MNRAS.461L.122L, 2017JCAP...03..023W, 2020MNRAS.491.1498C}).

(4) Our model predicts the repeating fraction $f_{\rm rep}(<z)$ for sources within redshift $z$, which depends on the distribution of activity levels among different FRB sources and is generally a decreasing function of redshift, as shown in Figures \ref{fig:frep_tot} and \ref{fig:frep_z}. This can be applied once we know the $\dmex$ distributions of both repeaters and the apparent non-repeaters in the CHIME sample. For instance, if most sources are similar to FRB 121102, then we predict (i) more than 10\% of all CHIME sources should be repeaters, and (ii) at sufficiently low redshifts $z\lesssim 0.1$ (or $\dmex\lesssim 100\rm\, pc\,cm^{-3}$) nearly all sources should be observed as repeaters by CHIME. Violation of either of them means that most sources are repeating much less frequently than FRB 121102.

\section{Data Availability}
The data underlying this article will be shared on reasonable request to the corresponding author.

\section{acknowledgments}
We thank Paz Beniamini for useful discussions. We thank the organizers for the 2019 Toronto FRB Day workshop and the 2020 Flatiron Institute FRB workshop where the current work was initiated. WL is supported by the David and Ellen Lee Fellowship at Caltech.

{\small
\bibliographystyle{mnras}
\bibliography{refs}
}

\appendix

\section{Weibull waiting time distribution for a single repeater}\label{sec:single_repeater}
To study the distribution of bursts from a repeating source, we use the Weibull probability density function (PDF) which has been used to model bursts from FRB 121102 \citep{2018MNRAS.475.5109O}. For a time interval $\delta$ between adjacent bursts of isotropic energy above $E$, the PDF is
\begin{equation}
  \label{eq:1}
  W(\delta;\lambda(E), k) = \lambda k (\lambda \delta)^{k-1}
  \mr{exp}[-(\lambda \delta)^k].
\end{equation}
where $\lambda(E)$ is related to the mean repeating rate above energy $E$ and $k$ is a shape parameter. The cumulative density function (CDF) and mean interval are given by
\begin{equation}
  \label{eq:2}
  \mr{CDF}(\delta) = \int_0^\delta W(\delta) \d
\delta  = 1 - \mr{exp}[-(\lambda \delta)^k],
\end{equation}
\begin{equation}
    \lara{\delta} =\int_0^\infty \delta \,W(\delta) \d
\delta = r^{-1},\ \ r \equiv {\lambda\over \Gamma(1+1/k)},
\end{equation}
where $\Gamma(x)$ is the gamma function and $r$ is the mean repeating rate. 

\begin{figure*}
\centering
\includegraphics[width=0.85\textwidth]{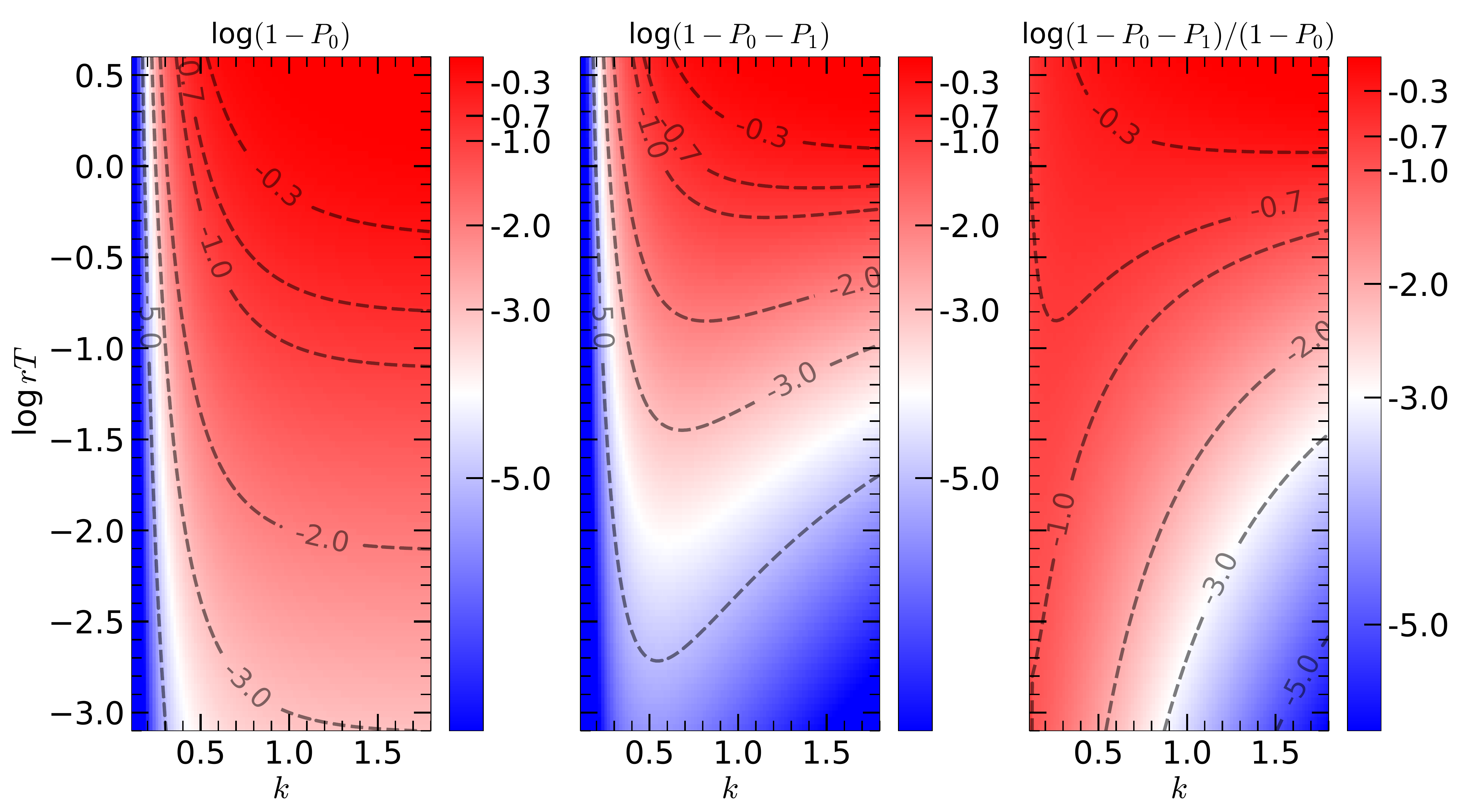}
\caption{Probabilities of detecting at least one (left) or at least two (middle) bursts, and the ratio of these two probabilities (right panel), for a single observing session of duration $T$. In the limit of small Weibull shape parameter $k\ll 1$ (the highly clustered case), the detection probability $1-P_0$ is typically much less than the Poissonian value of $rT$. Another consequence of strong temporal clustering ($k\ll 1$) is that the repeating fraction $(1-P_0-P_1)/(1-P_0)$ is much higher than the corresponding Poissonian case.
}
\label{fig:single_obs}
\end{figure*}

\begin{figure*}
\centering
\includegraphics[width=0.85\textwidth]{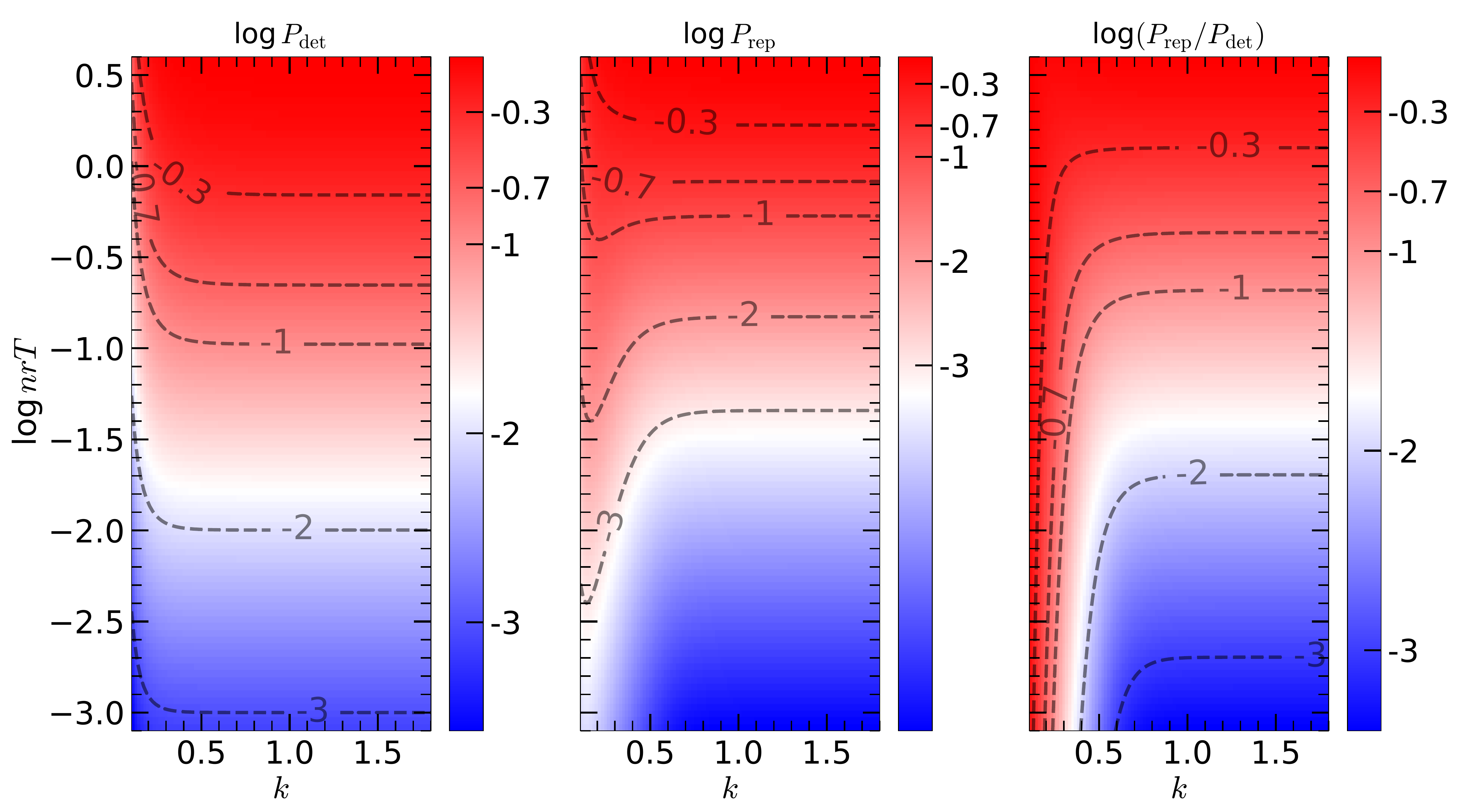}
\caption{Probabilities of detecting at least one (left) or at least two (middle) bursts, and the ratio of these two probabilities (right panel), for $n=150$ independent identical observing sessions of duration $T$. With a large number of observing runs, the detection probability $P_{\rm det}$ is close to the Poissonian value $nrT$ for $k\gtrsim 1$ and is much higher for $k\ll 1$. Strong temporal clustering ($k\ll 1$) also increases the fraction of repeating bursts out of all the detected ones, $P_{\rm rep}/P_{\rm det}$.
}
\label{fig:mul_obs}
\end{figure*}

For a single continuous observing run lasing for $T$, one can derive that the probability of seeing zero events is
\begin{equation}
  \label{eq:3}
  P_0(\lambda, k) = {1\over k\Gamma(1 + 1/k)} \Gamma_{\rm in}(1/k, (\lambda
  T)^k),
\end{equation}
the probability of seeing at least one event is
\begin{equation}
    1 - P_0(\lambda, k) = {1\over k\Gamma(1 + 1/k)} \gamma_{\rm in}(1/k,
  (\lambda T)^k), 
\end{equation}
and the probability of seeing exactly one event is
\begin{equation}
  \label{eq:4}
  P_1(\lambda, k) = {\lambda T \over \Gamma(1+1/k)} \int_0^1
  \mr{exp}\left[-(\lambda T)^k(x^k + (1-x)^k)\right] \d x. 
\end{equation}
In these expressions, $\Gamma_{\rm in}(s, x) = \int_x^\infty t^{s-1} \mr{e}^{-t}\d t$ and $\gamma_{\rm in}(s, x) = \int_0^x t^{s-1} \mr{e}^{-t}\d t$ are the upper and lower incomplete gamma functions respectively. These probabilities are shown in Figure \ref{fig:single_obs}. In the limit $(\lambda T)^k\ll 1$, the lowest order expansion of $\mr{exp}(-x)\approx 1-x$ gives
\begin{equation}
  \label{eq:9}
  1-P_0\approx r T \left[1 - {(\lambda T)^k
    \over k+1}\right],
\end{equation}
\begin{equation}
    P_1\approx r T \left[1 - {2(\lambda T)^k
    \over k+1}\right],
\end{equation}
and the probability of detecting two or more bursts is
\begin{equation}
  \label{eq:10}
  {1 - P_0 - P_1} \approx rT{(\lambda T)^k \over k+1}.
\end{equation}
Therefore, for a \textit{single} observing session, if the chance of detection is small $1-P_0\simeq r T\ll 1$, the probability of identifying the source as a repeater is even smaller by another factor of $(\lambda T)^k/(k + 1)$.

If there are $n$ independent observing runs of identical durations $T$ (so that the total duration is $nT$), the probability of detecting at least one burst is
\begin{equation}
  \label{eq:6}
  P_{\rm det}(\lambda, k) = 1 - P_0^n.
\end{equation}
The probability of single detection is
\begin{equation}
  \label{eq:13}
  P_{\rm sig}(\lambda, k) = nP_1P_0^{n-1}.
\end{equation}
The probability of repetitive detection (of at least two events) is
\begin{equation}
  \label{eq:7}
  P_{\rm rep}(\lambda, k) = P_{\rm det} - nP_1P_0^{n-1}.
\end{equation}
The fraction of repetition is
\begin{equation}
  \label{eq:8}
  f_{\rm rep} = {P_{\rm rep}\over P_{\rm det}} = {1-P_0^n(1 + nP_1/P_0) \over 1-P_0^n}.
\end{equation}
These probabilities are shown in Figure \ref{fig:mul_obs}. The formalism can be generalized into $n$ runs of non-equal durations, but the set-up of identical sessions is reasonable since the CHIME beams sweep across the position of each source on a regular basis. Our numerical results are based on the exact expressions above.

For the purpose of intuitive understanding, we consider the limit $nr T\ll 1$ or $P_{\rm det}\ll 1$ (such that simultaneous observation of a large number $P_{\rm det}^{-1}\gg 1$ of sources is needed to yield a detection), and the probabilities can be simplified to the first order
\begin{equation}
  \label{eq:11}
  P_{\rm det} \approx nrT\left[
1 - {(\lambda T)^k \over k+1} - {n-1\over 2} rT
\right],
\end{equation}
\begin{equation}
  \label{eq:12}
  P_{\rm sig} \approx nrT \left[
1 - 2{(\lambda T)^k \over k+1} - (n-1) rT
\right],
\end{equation}
\begin{equation}
  \label{eq:14}
  P_{\rm rep}\approx nrT\left[{(\lambda T)^k \over k+1} + {n-1\over 2}
    rT\right]. 
\end{equation}
To the zeroth order, $P_{\rm det}\simeq nrT$ means that the observing time can be linearly added independent of the duration of each session and that the mean occurrence rate is $r \equiv \lambda/\Gamma(1+1/k)$. This applies to the Commensal Real-time ASKAP Fast Transient (CRAFT) survey because the telescope is only sensitive to the brightest and rarest bursts from each source \citep{2018Natur.562..386S} and hence $r$ is very small. Fixing $\lambda$, $k$ and $T$, we see that the fraction of repetition increases as more and more sessions of observations are carried out.

\section{The stochastic relation between redshift and $\dmex$}\label{sec:DM_z}

An important aspect of comparing our models with the CHIME observations is converting from $N_{\rm rep}(<z)$ to $\d N /\rm \d log DM_{\rm ex}$. We describe our approach to this in more detail below.

The free electrons within the host galaxy (ISM and halo) and in the IGM along the line of sight both contribute to $\dmex$ beyond the Milky Way. Each of these two components may have stochastic fluctuations, depending on the host galaxy properties and the number of intervening halos. For a given source at redshift $z$, we approximate these two components as random Gaussian variables with mean and standard deviation: $\mu_{\rm host}$, $\sigma_{\rm host}$; $\mu_{\rm IGM}$, $\sigma_{\rm IGM}$. Then, the PDF of $\dmex$ is given by
\begin{equation}
\begin{split}
    {\d P\over \d \dmex} & (\dmex, z)  = {1\over \sqrt{2\pi (\sigma_{\rm host}^2 + \sigma_{\rm IGM}^2)}}\\
    &\, \times \mr{exp}\left[-{(\dmex - \mu_{\rm host} - \mu_{\rm IGM})^2 \over 2 (\sigma_{\rm host}^2 + \sigma_{\rm IGM}^2)}\right].
\end{split}
\end{equation}
We adopt $\mu_{\rm host} = 100/(1+z)$, $\sigma_{\rm host} = 30/(1+z)$, $\mu_{\rm IGM}(z) = 900z$, $\sigma_{\rm IGM} = 200\sqrt{z}$, all in units of $\rm\, pc\,cm^{-3}$. These values are motivated by observational and theoretical studies of (potential) FRB host galaxies \citep{2017ApJ...834L...7T, 2019Sci...365..565B, 2019Natur.572..352R, 2019Sci...366..231P, 2020Natur.577..190M, 2015RAA....15.1629X, 2018MNRAS.481.2320L} and IGM electron density distribution \citep{2014ApJ...780L..33M, 2014ApJ...783L..35D, 2018ApJ...852L..11S, 2019ApJ...872...88R, 2019MNRAS.485..648P, 2019PhRvD.100h3533K}. The $\sigma_{\rm IGM}\propto \sqrt{z}$ behavior can be roughly understood if one divides the line of sight into many short segments (of e.g., $\sim 50\rm\, Mpc$) each of which has fractional DM fluctuation of order unity due to (on average) one intervening massive halo, so the sum of $n$ segments gives fractional fluctuation of $n^{-1/2}$. In reality, the $\rm DM_{\rm IGM}$ fluctuation is non-Gaussian with a long tail at high DM given by non-zero probability of an intervening galaxy cluster. Our analysis is only weakly affected by the choice of the above parameters, which may change as our understanding of FRB host galaxies improves.

The above normal distribution allows (unphysical) negative $\dmex$ but at a very low probability, which we ignore. Then the following convolution
\begin{equation}
    {\d N_{\rm rep}\over \d \dmex} (\dmex) = \int_0^\infty \d z {\d N_{\rm rep}\over \d z}(z) {\d P\over \d \dmex} (\dmex, z).
\end{equation}
provides the relation between $\d N_{\rm rep}/\d z$ in equation~(\ref{eq:41}) and the desired distribution of $\d N_{\rm rep}/\d \dmex$.







\section{Correlation between observing sessions}\label{sec:correlation}
Our calculations are based on the simplified assumption that observing sessions are independent, which makes the probability of singly or multiply detections analytically tractable as given in Appendix \ref{sec:single_repeater}. In Fig. \ref{fig:Prep_comparison}, we compare the repeating probability $P_{\rm rep}$ from our simplified approach and that from direct sampling of the Weibull distribution across all observing sessions (maintaining the full correlation). We find that the repeating probability given by eq. (\ref{eq:7}) is generally different from that from direct Monte Carlo sampling. However, the difference between the two methods can be effectively reconciled by shifting to a different $k$ (e.g., $k=0.3\rightarrow0.2$). Therefore, our final results, after marginalizing over $k$, should not be strongly affected by the assumption of independent observing sessions. We also note that, given our insufficient understanding of the true wait-time distribution, it is unclear if using the full Weibull distribution across the entire observing run is closer to reality than our simple method based on uncorrelated observing sessions. This limitation of our model can be improved when we have a better understanding of the wait-time distribution of repeating FRBs (by monitoring a large number of them) in the future.

\begin{figure}
\centering
\includegraphics[width=0.45\textwidth]{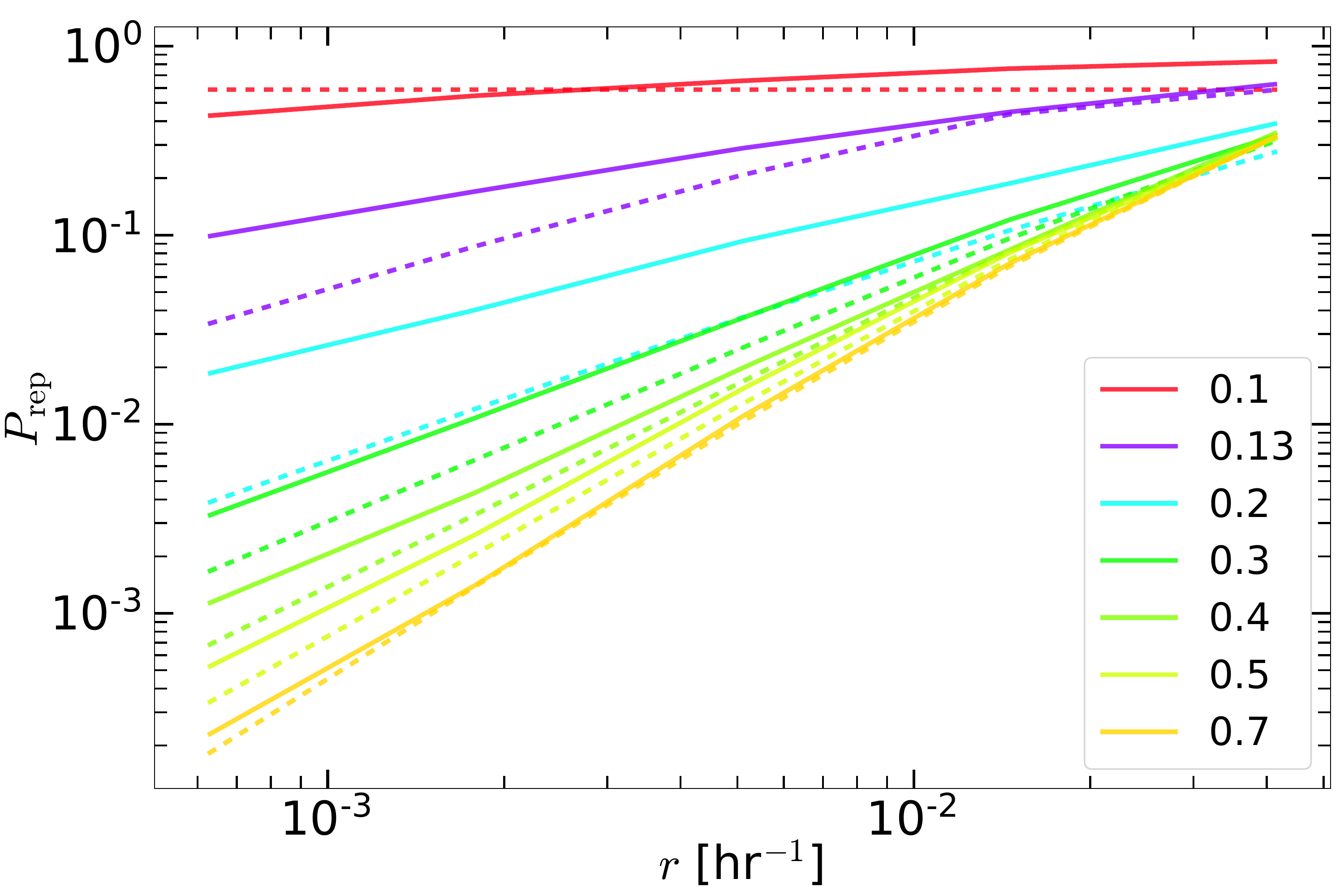}
\caption{The probability of having two or more detections $P_{\rm rep}$ as a function of repeating rate $r$ for different shape parameters $k$ (color-coded) shown in the legend. Here we consider $n=150$ daily observing sessions each lasting for $T=0.2\rm \, hr$. The solid lines are from direct sampling of the Weibull distribution across the entire 150 day observing run and bursts occurring within the 0.2 hour on-source time are recorded as detections. For each $k$ and $r$, we simulate a large number ($\sim10^5$) of random cases so that the repeating probability $P_{\rm rep}$ is well converged. The dashed lines are from eq. (\ref{eq:7}) assuming that each of the $n$ sessions are independent of each other.
}
\label{fig:Prep_comparison}
\end{figure}

\label{lastpage}
\end{document}